# Weak itinerant ferromagnetism in Heusler type Fe$_2$VAl$_{0.95}$

K. Sato,[1] T. Naka,[1,*] M. Taguchi,[1] T. Nakane,[1] F. Ishikawa,[2] Yuh Yamada,[3] Y. Takaesu,[4] T. Nakama,[4] A. de Visser,[5] and A. Matsushita[1]

[1] *National Institute of Materials Science, Tsukuba, Sengen 1-2-1, Tsukuba, Ibaraki 305-0047, Japan*

[2] *Graduate School of Science and Technology, Niigata Univ. 8050 , Igarashi Niigata, 950-2181, Japan*

[3] *Department of Physics, Faculty of Science, Niigata Univ. 8050 , Igarashi Niigata, 950-2181, Japan*

[4] *Faculty of Science, University of the Ryukyus, Nishihara, Okinawa 903-0213, Japan*

[5] *Van der Waals –Zeeman Institute, University of Amsterdam, Valckenierstraat 65, Amsterdam 1018 XE, The Netherlands*

We report measurements of the magnetic, transport and thermal properties of the Heusler type compound Fe$_2$VAl$_{0.95}$. We show that while stoichiometric Fe$_2$VAl is a non-magnetic semi-metal a 5% substitution on the Al-site with the 3$d$ elements Fe and V atoms leads to a ferromagnetic ground state with a Curie temperature $T_C$ = 33±3 K and a small ordered moment $\mu_s$ = 0.12 $\mu_B$/Fe in Fe$_2$VAl$_{0.95}$. The reduced value of the ratio $\mu_s/\mu_p$ = 0.08, where $\mu_p$ = 1.4 $\mu_B$/Fe is the effective Curie-Weiss moment, together with the analysis of the magnetization data $M(H,T)$, show magnetism is of itinerant nature. The specific heat shows an unusual temperature variation at low temperatures with an enhanced Sommerfeld coefficient, $\gamma$ = 12 mJK$^{-2}$mol$^{-1}$. The resistivity, $\rho(T)$, is metallic and follows a power law behavior $\rho = \rho_0+AT^n$ with $n \approx 1.5$ below $T_C$. With applying pressure, $T_C$ decreases with the rate of $(1/T_C)(dT_C/dP)$ = -0.061 GPa$^{-1}$. We conclude substitution on the Al-site with Fe and V atoms results in itinerant ferromagnetism with a low carrier density.

Key words: itinerant ferromagnetism, specific heat, resistivity, magnetization, Heusler compound

PACS: 75.30.-m; 75.30.Kz; 75.47.-m; 71.20.Be

*Corresponding author.
Tel.: +81-22-217-5630; Fax: +81-22-217-5631; E-mail: naka.takashi@nims.go.jp

## I. Introduction

Weak ferromagnetic (or antiferromagnetic) transition metal and rare earth or actinide metallic compounds are intensively investigated, since novel ground states, like heavy fermion behavior, non-Fermi liquid states and unconventional superconductivity, may emerge at low temperatures. These exotic phases are located at the boundary of magnetically ordered states, and may be reached by tuning the ground state by chemical pressure (alloying), mechanical pressure and/or applying an external magnetic field. Heusler and semi-Heusler type of compounds composed of transition metals form an attractive materials family to investigate novel emerging phenomena, the more because these are candidates for half metallic states, in which the band electrons are fully spin-polarized [1-2]. In the magnetically ordered state of half metals, the exchange gap is quite large and the minority spin band is fully occupied. This might be of practical use in spin sensitive devices [3], which provides a second reason why Heusler-type compounds are investigated intensively. Since applications demand both a high spin polarization and stable ferromagnetism above room temperature, this is an exigent and challenging mission.

Although the concept of the half-metal comes truly from an itinerant electron picture, one can address the question how much local moment character strongly spin polarized metals possess [4-5]. In fact, it is well known that in nearly and weak ferromagnetic itinerant electron systems, such as $ZrZn_2$ and MnSi, the conduction electrons are responsible for both the itinerant and localized electron nature (*i.e.* the Curie-Weiss behavior in magnetic susceptibility). For a few candidate materials only the low-temperature magnetic and thermal properties could be analyzed satisfactorily in the framework of itinerant electron magnetism. We mention the semi-Heusler type ferromagnet NiMnSb [4] and antiferromagnet CuMnSb [5], which were examined under magnetic field and high pressure.

In this paper we focus on the magnetic transition and the low-temperature properties of the Heusler type iron-vanadium compound, $Fe_2VAl_{0.95}$. $Fe_2VAl$ and related alloys have attracted considerable attention [6-11], since it was claimed that these 3$d$-electron systems with small carrier concentration, $n \approx 0.01$ per unit formula [7], exhibit a significant carrier mass enhancement and non-Fermi liquid behavior [6]. Stoichiometric $Fe_2VAl$ is non-magnetic and related alloys, such as $Fe_{2+x}V_{1-x}Al$ and $Fe_2VAl_{1-\delta}$, exhibit ferromagnetic transitions [7, 10-14]. Consequently, the samples at $x \approx 0$ and $\delta \approx 0$ are located at the brink of ferromagnetic order, close to the ferromagnetic quantum critical point, *i.e.* $T_C \rightarrow 0$ [6,8]. Stimulated by these findings, theoretical work claimed $Fe_2VAl$ is a non-magnetic semimetal with the pseudogap at the Fermi level [15-17]. The considerable mass enhancement of the conduction carriers was attributed to excitonic correlations [16] or spin fluctuations [17]. However, subsequent specific heat, $C(T)$, measurements in applied magnetic fields showed that the upturn in $C/T$ at low temperatures results from a Schottky contribution of magnetic clusters in $Fe_2VAl$ [8]. The effects of off-stoichiometry have not been investigated quantitatively in the framework of nearly and weak ferromagnetism at $x \approx 0$ and $\delta \approx 0$ yet. Another interesting aspect of $Fe_2VAl$ is its potential for application in thermoelectric devices, since off-stoichiometry and heat-treatment procedures may results in an enhancement of the Seebeck coefficient [6, 11-12]. In $Fe_{2+x}V_{1-x}Al$ [11-12] and $Fe_2VAl_{1-\delta}$ [12], the thermoelectric power is enhanced for $x \approx 0$ and $\delta \approx 0$: with a negative sign for $x < 0$ and $\delta > 0$, and positive sign for $x > 0$ and $\delta < 0$. In the case of stoichiometric $Fe_2VAl$ the negative temperature coefficient in the electrical resistivity observed up to 1300 K and the constant value of the Hall coefficient at low temperatures [6-7, 12] suggest it is a low carrier-density semimetal with a pseudogap at the Fermi level [6]. Anti-site defects and magnetic nanoclusters induced by heat treatment and nominal off-stoichiometry bring about ferromagnetism [7,10,13,18] and superparamagnetism [18]. With increasing $x$ in $Fe_{2+x}V_{1-x}Al$, the excess of Fe atoms occupy the nominal V-site and, consequently, $Fe_3Al$ ($Fe_2FeAl$) clusters are formed (notice bulk $Fe_3Al$ is ferromagnetic with $T_C$=803 K [6, 13]). The presence of anti-site Fe atoms on the nominal V site in the $FeV_2Al$ system brings about a simultaneous enhancement of the magnetization and resistivity at low temperatures [7, 9-10]. This led to the claim that localized magnetic moments must be a crucial factor in the strong magnetic scattering of the conduction electrons responsible for the steep rise of the resistivity, the negative temperature coefficient of the resistivity, $d\rho/dT < 0$ and a giant magnetoresistance (GMR) at $x > 0$. Thus it is proposed that the steep rise of the electrical resistivity at lower temperature is not due to the energy gap but due to strong magnetic scattering [7, 9-10].

Al poor $Fe_2VAl_{1-\delta}$ ($\delta > 0$), seems to undergo a ferromagnetic transition and shows metallic conduction below room temperature [12]. The $Fe_2VAl_{1-\delta}$ system at $\delta > 0$ yields the opportunity to examine quantitatively not only the magnetic, but also the transport and thermal properties within the framework of itinerant electron ferromagnetism, as established for intermetallic compounds, like MnSi and $ZrZn_2$ [19]. In this paper we report a comprehensive study of the low temperature bulk properties of $Fe_2VAl_{0.95}$, notably magnetization, electrical resistivity and specific heat measurements. Whereas, as pointed out above, the magnetism in $Fe_{2+x}V_{1-x}Al$ is considered to be due to magnetic clusters having a semi-localized nature [8], our results on $Fe_2VAl_{0.95}$ show it is a weak ferromagnet, with a bulk magnetic transition at $T_C = 33$ K and a low carrier concentration $n \approx 0.06$.

## II. Experimental

Polycrystalline samples of $Fe_2VAl_{0.95}$, $Fe_2VAl$ and $Fe_2VAl_{1.05}$ were prepared by arc melting the proper amounts of elements into the nominal chemical composition. The samples were subjected to a heat treatment at 1000 °C for 15 hours for homogeneization purposes, and next annealed at 400 °C for 15 hours. For all samples the crystal structure was identified as the Heusler type. No secondary phases were detected in the powder X-ray diffraction (XRD) profiles. Since after annealing the grain size (*i.e.* the individual crystallite size) in the polycrystalline samples is quite large (> several hundredths μm), we expect that the volume fraction of impurity phases located at the grain boundaries is quite low. The samples we report on in this work are the same as those previously characterized by XRD, scanning electron microscope (SEM) and energy dispersive X-ray spectroscopy (EDS) [7].

Dc-magnetization, 4-probe ac electrical resistivity and specific heat measurements down to $T$ = 2 K were carried out using a conventional SQUID magnetometer (MPMS-XL) and a physical property measurement system (PPMS - Quantum Design), respectively. Resistivity measurements down to $T$ = 0.23 K were carried out in a $^3$He refrigerator (Heliox VL, Oxford Instruments) with a sensitive LR700 (Linear Research) ac resistance bridge at the Van der Waals-Zeeman Institute of the University of Amsterdam. In the high pressure measurement, a handmade measurement system specialized for high pressure measurement with a glass dewar vessel was used. To generate high pressure a hybrid clamp cell made from nonmagnetic NiCrAl and CuBe alloys was used. The sample was mounted on a specially designed plug and put inside a Teflon cell with the pressure-transmitting medium Daphne 7373 (Idemitsu).

## III. Results and discussion

### A. Crystallographic

X-ray diffraction patterns of $Fe_2VAl_{1-\delta}$ at $\delta$ = -0.05, 0, and 0.05 are apparently identified to be of the single phase Heusler type ($Cu_2MnAl$) structure in the space group Fm3m. A as shown in Fig. 1(a), no impurity phases were detected, except for small amounts of Al and Al oxide inclusions indicated by SEM and EDX measurements [7]. Therefore, we expect, the actual Al composition, $\delta$, is slightly smaller than the nominal one. Figure 1(b) shows the $\delta$-dependence of lattice constant, $a$, estimated in this work with those reported in previous works [18, 20] for comparison. The lattice constant at $\delta = 0$ measured in this work corresponds well with those for annealed samples. The linear dependence of $a$ versus $\delta$ supported that the excess Fe and V atoms occupy the Al site for $\delta > 0$ and the excess Al atom locates at the Fe and/or V site(s) for $\delta < 0$, since Al has smaller ionic radius than those of Fe and V.

As shown below (see section IIIC), the resistivity *versus* temperature curve obtained for $\delta = 0.05$ is quite similar to those reported by Nishino *et al*. [12]. Rounded maxima in the resistivity were observed at $T_{max}^{low}$ = 40 K and $T_{max}^{high}$ = 250 K in a sample with composition $(Fe_{2/3}V_{1/3})_{75.7}Al_{24.3}$ [12], which corresponds to $\delta = 0.035$ in the notation $Fe_2VAl_{1-\delta}$ used here. The maximum at $T = 40$ K in the resistivity for $\delta = 0.035$ locates the Curie temperature slightly lower than $T_{max}^{low}$ = 50 K of $\delta = 0.05$ observed in this work (Fig. 5(a)). Following the argument of the effect of off-stoichiometry based on the rigid band model [12, 21] which is consistent with the Al content variations of the sign in Seebeck and Hall coefficients [7, 12], we are convinced that the excess Fe and V atoms in $Fe_2VAl_{1-\delta}$ occupy the Al site for $\delta > 0$, and vice versa. Actually, it was shown recently by comprehensive magnetic and transport studies in the vicinity of $\delta \approx 0$ that not only the Curie point, $T_C$, but also the saturation moment, $\mu_s$, shows a continuous variation as a function of $\delta$ [22].

**B. Magnetization**

In Fig. 2(a) we show the magnetization of $Fe_2VAl_{0.95}$ as a function of magnetic field, $M(H)$, measured at various fixed temperatures. A spontaneous moment appears in the temperature range $T$ = 30-40 K, which indicates ferromagnetic order (see Fig. 4 (a)). In Fig. 2(b) we have traced the temperature variation of the magnetic susceptibility, $\chi(T)$, measured in a field of 20 kOe. As indicated by the straight line in the plot of $1/\chi(T)$ *versus* $T$ the susceptibility follows a Curie-Weiss law, $\chi(T) = C/(T-\theta)$, for $T$ > 100 K, from which we extract the Curie constant $C$ = 0.243 emuK/g and a Curie-Weiss temperature $\theta$ = 51 K. By using $C = N\mu_p^2\mu_B^2/3k_B$, an effective magnetic moment $\mu_p$ = 1.4 $\mu_B$/Fe is obtained, where $N$, $\mu_B$ and $k_B$ are the number of magnetic (Fe) atoms, the Bohr magneton and Boltzmann's constant, respectively. This value of $\mu_p$ is considerably larger than that of heat-treated $Fe_2VAl$ ($\mu_p$ ~ 0.2 $\mu_B$/Fe) [23]. The saturation moment at $T$ = 2 K is $\mu_s$ = 0.12 $\mu_B$/Fe, which is much smaller than the paramagnetic moment, $\mu_p$ = 1.4 $\mu_B$/Fe. The small value of the ratio $\mu_s/\mu_p$ = 0.083 corroborates the itinerant electron nature of ferromagnetism in $Fe_2VAl_{0.95}$. It is noteworthy that an Al-site substitution of only 5% ($\delta$ = 0.05) modifies the magnetic properties strongly: the appearance of a spontaneous moment and low magnetic transition temperature and a strong enhancement of the paramagnetic moment. We stress that these changes cannot be due to an impurity phase, like for instance $Fe_{1-x}V_x$ ($x$=1/3), for which the estimated values of $\mu_p$ and $\mu_s$ are one order of magnitude larger than those observed and $T_C$ is above room temperature [24]. In contrast with the observations concerning magnetism in annealed $Fe_2VAl$ [18] and $Fe_2V_{1-x}Cr_xAl$ [25], the ferromagnetic transition in $Fe_2VAl_{0.95}$ is also signaled: (*i*) in the temperature derivative of the resistivity, $(1/\rho)(d\rho/dT)$, shown in Fig. 5(b), (*ii*) by the temperature variation of the specific heat divided by temperature, $C/T$, showing a lambda-peak-like shape (Fig. 9(b)), (*iii*) by the $M^4$-$M/H$ plot (Fig. 3(a)), and (*iv*) by the $M^2$ $-(T/T_C)^{4/3}$ plot (Fig. 4 (a)). Notably, the transition temperatures $T_C=33\pm3$ K determined by the different techniques agree within the uncertainty of 3 K. Correspondingly, both coercive force, $H_c$, and remanent magnetization, $M_r$, emerge around $T_C$ (Fig. 2(b)).

As mentioned in the Introduction, it is interesting to examine whether magnetism in Fe$_2$VAl$_{0.95}$ has an itinerant or localized character. In the latter case it is expected that the induced spin is localized at the 3$d$ atoms (Fe or V) substituted on the Al site. In Figs 3(a) and 3(b) we analyze the magnetization data around $T_C$ in terms of $M^4$-$H/M$ and $M^2$-$H/M$ (Arrott) plots, respectively. For weak ferromagnetic systems the $M^4$ versus $H/M$ curve at $T = T_C$ is predicted to be linear with an intercept zero [26]. Inspecting Fig. 2(a) we conclude $T_C = 33$ K. Notice, in the Arrott plot the $M^2$ versus $H/M$ curves all exhibit a strong curvature towards the $H/M$-axis. This feature is quite similar to those found in prototypical itinerant ferromagnets, like MnSi and Ni$_3$Al [26]. Thus the magnetization data provide strong evidence for itinerant ferromagnetism with $T_C = 33$ K in Fe$_2$VAl$_{0.95}$. Based on the generalized Rhodes-Wohlfarth plot [26], we obtain the characteristic temperature $T_0 = 895$ K for Fe$_2$VAl$_{0.95}$. The ratio $T_C/T_0$ is an important parameter as it characterizes the degree of localization or itineracy of the spin moment [19]. For $T_C/T_0 \ll 1$ the magnetic material has a strong itinerant character, while at $T_C \sim T_0$ local moment magnetism results. The estimated value $T_C/T_0 = 0.038$ is comparable to the values reported for MnSi (0.13), Ni$_3$Al (0.0116), ZrZn$_2$ (0.053) and Sc$_3$In (0.0097) which all exhibit weak ferromagnetism [26]. With help of $T_0$, one may calculate the temperature dependence of the magnetization below $T_C$. With decreasing $T_C/T_0$ the $M(T)$ curve deviates more and more from the mean field behavior, $M = M(0)(1-(T/T_C)^2)^{1/2}$, and has a tendency to follow $M = M(0)(1-(T/T_C)^{4/3})^{1/2}$. In fact, as shown in Fig. 4(b), $M^2(T)$ of Fe$_2$VAl$_{0.95}$ obeys the $T^{4/3}$-dependence for a weak itinerant ferromagnet rather than the $T^2$-dependence of the mean field model [26]. By using $M(T) =$ $M(0)(1-(T/T_C)^{4/3})^{1/2}$ the zero temperature magnetization and Curie point are estimated to be $M(0) = 5.7$ emu/g and $T_C = 36$ K at $H = 200$ Oe, respectively.

## C. Electrical resistivity

In Fig. 5(a) we show the temperature dependence of the resistivity, $\rho(T)$. The resistivity, $\rho(T)$, exhibits rounded maxima at $T_{max}^{low} = 52$ K and $T_{max}^{high} = 260$ K. A comparison with $Fe_{2+x}V_{1-x}Al$ at $x \approx 0$ [7, 10, 14] reveals that the resistivity above $T \sim 260$ K is dominated by scattering due to thermal excitations in the pseudo gap. Below $T \sim 260$ K $\rho(T)$ exhibits metallic behavior, with a low carrier number [7]. On the other hand, the maximum at $T_{max}^{low} = 52$ K is caused by the para-to-ferromagnetic phase transition, as was also observed in the Fe-rich composition of $Fe_{2+x}V_{1-x}Al$ [7, 10, 14]. In $Fe_2VAl_{0.95}$ $\rho(T)$ remains metallic down to $T \sim 1$ K, while in the case of $Fe_{2+x}V_{1-x}Al$, especially for $x \leq 0.05$, a semiconductor behavior is reported. The latter was corroborated by weak localization effects of the carriers in a random potential [9]. As shown in Fig. 5(b), the sharp anomaly in $(1/\rho)(d\rho/dT)$ is indicative of a bulk ferromagnetic transition at $T = 33$ K. In the ferromagnetic phase the resistivity follows a power law behavior $\rho(T) \sim T^n$ with $n \cong 3/2$ (see inset to Fig. 5(a)). Notice, it was derived by the SCR theory that the temperature dependence of $\rho$ should follow a power law $\sim AT^2$ at $T<T_C$ [19].

## D. Magnetoresistance and low temperature electrical resistivity

The itinerant ferromagnetic character of $Fe_2VAl_{0.95}$ is also reflected in the transverse magnetoresistance (TMR), $(\rho(H)-\rho(0))/\rho(0)$, especially around $T = T_C$, as shown in Fig. 6(a).

The TMR is negative below 100 K and has a minimum at a temperature slightly above $T_C$. The TMR reaches a maximum size of -11% at $T = 43$ K in $H = 80$ kOe. The negative temperature coefficient between 50 K $< T <$ 100 K and the maximum in $\rho(T)$ slightly above $T_C$ are wiped out above $H = 30$ kOe, as shown in Fig. 6(a). According to a numerical calculation based on the SCR theory [27], the negative magnetoresistance is due to the presence of spin fluctuations above $T_C$ and, consequently, the maximum temperature in the TMR shifts toward higher temperature with increasing magnetic field. This field dependence of TMR is comparable with that in the itinerant ferromagnet $Sc_3In$ ($T_C = 5.3$ K) [27-28]. For $T > 100$ K, the TMR is positive. These transport data differ from those in non-magnetic $Fe_{1.98}V_{1.02}Al$ [7], where $\rho(T)$ exhibits a rounded maximum around room temperature, and the magnetoresistance is negative. Remarkably, the temperature dependence of the resistivity changes to $\rho = \rho_0 + AT^2$ below a certain temperature $T^*$, which increases magnetic field (Fig. 6(b)). This feature is quite similar to that of non-Fermi liquid systems, such as $U_3Ni_3Sn_4$ where the Fermi liquid behavior is recovered under magnetic field or external pressure [29].

**E. Pressure effects on resistivity**

As mentioned above, crystallographic and chemical defects affect strongly on the magnetic grand state of $Fe_2VAl$ and related alloys [7]. Remarkably, lattice constant of $Fe_2VAl$ reflects sensitively these structural deviations from the Heusler structure [18]. Plausibly, Curie temperature, $T_C$, is a function of not only carrier number but also the variation of lattice constant, $\Delta a/a$. To distinguish those effects on the ferromagnetic transition in $Fe_2VAl_{0.95}$, it is expected that resistivity measurement under high pressure provides crucial information. The lattice expansion, $\Delta a/a = 0.10$ % at $\delta = 0.05$, can be compensated by applying an experimentally accessible pressure, $\Delta P = 3(\Delta a/a)_{\delta=0.05}/\kappa_V \sim 0.5$ GPa where $\kappa_V$ is volume compressibility estimated to be $6.81 \times 10^{-3}$ GPa$^{-1}$ for $\delta = 0$ [20]. As shown in the inset of Fig. 7, $T_{max}^{high}$ increases while $T_{max}^{low}$ decreases with pressure. The increment of $T_{max}^{high}$ with applying pressure can be responsible for that the pseudo gap locating at the Fermi level increases. On the other hand, the decrement of $T_{max}^{low}$ seems to be related to that of $T_C$. Pressure dependence of $T_C$ obtained from the temperature variation of $(1/\rho)(d\rho/dT)$ is shown in Figs. 8(a) and (b). With increasing pressure the anomalous point shifts toward lower temperature with the pressure coefficient of $(1/T_C)(dT_C/dP) = -0.061$ GPa$^{-1}$ (Fig. 8 (b)), which corresponds well with $(1/T_{max}^{low})(dT_{max}^{low}/dP) = -0.074$ GPa$^{-1}$. The volume effect on $T_C$ is comparable with those for itinerant ferromagnets, such as Ni$_3$Al ($\sim -0.10$ GPa$^{-1}$ [30, 31]). Therefore, the emergence of the ferromagnetism at $\delta > 0$ can results from mainly the carrier doping in Fe$_2$VAl$_{1-\delta}$, while the volume expansion seems to be a miner effect. At $T << T_C$, it is noteworthy that $\rho_0$ increases with the pressure coefficient of $(1/\rho_0)(d\rho_0/dP) = 0.038$ GPa$^{-1}$ while the exponent of temperature, $n$, decreases from $n = 1.42$ at 0.11 GPa to 1.36 at 1.96 GPa, which obtained in a temperature range of $3.5<T<11$ K (Fig. 7).

**F. Specific heat**

In Fig.9(a) we show the specific heat divided by temperature, $C/T$, as a function of $T^2$ measured on Fe$_2$VAl$_{1-\delta}$ samples with $\delta = -0.05$, 0 and $+0.05$. A low-temperature upturn as present for Fe$_2$VAl$_{1.05}$ and Fe$_2$VAl was also observed previously in Fe$_2$VAl [6, 8] and in Fe$_{2-x}$V$_{1+x}$Al [10]. Remarkably, for Fe$_2$VAl$_{0.95}$ $C/T$ has a strong curvature towards the $T^2$ axis. Assuming that this curvature in $C/T$ is due to a ferromagnetic spin-wave contribution, $C_{sw} \sim aT^{3/2}$, we fitted the experimental curve at $H = 0$ to $C/T = \gamma + a_{sw}T^{1/2} + \beta T^2$ below 20 K (fit not shown in Fig. 9(a)). The fitting parameters are $\gamma = 12.6$ mJK$^{-2}$mol$^{-1}$, $a_{sw} = 1.38$ mJK$^{-5/2}$mol$^{-1}$ and $\beta = 0.025$ mJK$^{-4}$mol$^{-1}$. However, the presence of the spin-wave contribution in the low-temperature specific heat is not supported by measurements in magnetic field. In field the curvature in $C/T$ enhances significantly. At $H = 80$ kOe our fitting procedure results in a larger value for $a_{sw}$, whereas the spin wave contribution should be suppressed under magnetic field, as for instance in Ni [32] and CeRu$_2$Ge$_2$ [33]. This strongly suggests that the low temperature variation in $C(T)/T$ is not due to a ferromagnetic spin-wave contribution.

As indicated in Fig. 9(a), the difference in specific heat between $\delta = 0$ and -0.05 is quite small except in the low temperature region where $C/T$ showing an upturn. We fitted the high temperature specific heat, $C_{HT}$, in the temperature range 8-30 K to $C_{HT} = \gamma T + \beta T^3 + \eta T^5$ and estimated $\gamma = 3.10 \pm 0.02$ mJmol$^{-1}$K$^{-2}$, $\beta = 0.0265 \pm 0.0001$ mJmol$^{-1}$K$^{-4}$ and $\eta = (1.82 \pm 0.02) \times 10^{-5}$ mJmol$^{-1}$K$^{-6}$ for $\delta = -0.05$ and $\gamma = 2.19 \pm 0.02$ mJmol$^{-1}$K$^{-2}$, $\beta = 0.0283 \pm 0.0001$ mJmol$^{-1}$K$^{-4}$ and $\eta = (1.75 \pm 0.01) \times 10^{-5}$ mJmol$^{-1}$K$^{-6}$ for $\delta = 0$. These estimated parameters correspond well to those of Fe$_2$VAl obtained previously in the temperature range 8-25 K [8].

### G. Comparison with paramagnetic Fe$_2$VAl in the specific heat

Contrasted to the case of $\delta = -0.05$, $C/T$ of $\delta = 0.05$ is enhanced obviously compared with that of $\delta = 0$. Employing paramagnetic Fe$_2$VAl as a reference material in the specific heat, the compositional change in $C(T)$, $\Delta C = C(\text{Fe}_2\text{VAl}_{0.95}) - C_{HT}(\text{Fe}_2\text{VAl})$ is obtained as shown in Fig. 9(b). Obviously, at zero field a pronounced anomaly is present at $T \approx 34$ K, *i.e.* at the same temperature where $(1/\rho)(d\rho/dT)$ exhibits a maximum. This anomaly is superposed on the background which has a broad maximum around $T \sim 30$ K (see inset Fig. 9 (b)). The magnetic entropy is obtained by integrating $C_{FM}/T$ versus $T$ and is estimated at 0.7% of $R\ln2$, where $R$ and $C_{FM}$ are the gas constant and the ferromagnetic component in the specific heat, respectively. Such a small value is expected for a weak itinerant ferromagnet [19]. Below the phase transition point the $\Delta C/T$ curve decreases further with decreasing temperature. Below $T \sim 10$ K the specific heat can be expressed by the sum of linear ($\sim T$) and quadratic ($\sim T^2$) terms. The magnetic field wipes out the magnetic transition anomaly in $\Delta C/T$ and causes $\Delta C/T$ to decrease, as shown in Fig. 9(b). An anomalous $T^2$ temperature dependence in the specific heat was observed also in $Pd_2MnIn$ [34], GdCu [35], CuMnSb [36] and CeNiSn [37]. It seems to be due to a sharp density of state (DOS) shape around $E_F$, such as the V-shape density of states with small residual component at the Fermi level, as observed in the Kondo insulator CeNiSn [37]. In CeNiSn, at lower temperature the temperature variation of the specific heat shows a deviation from the $T$-linear dependence, which can be fitted by the sum of a $T$-linear and a $T^2$ terms [38]. Additionally, the band structure below $T_C$ in $Fe_2VAl_{0.95}$ is more complicated because the ferromagnetic ordering seems to modify considerably the DOS structure due to the exchange splitting, $\Delta_e$, between the spin-up and spin-down bands, which increases effectively with applying external field below $T_C$. In fact, the low temperature specific heat $\Delta C \sim \gamma T + \delta T^2$ changes, that is, the ratio of $\gamma/\delta$ decreases upon application of a magnetic field (Fig. 9(b)).

Next, let us discuss quantitatively the comparison between the magnetic moment, $\mu_p$, and carrier concentration, $n$, modified by the off-stoichiometry in $Fe_2VAl_{0.95}$. As shown in Fig. 6(b), the resistivity follows a quadratic temperature dependence below $T = T^*$, with $T^*$ increasing with magnetic field. The larger value of the coefficient of the $T^2$ term, $A = 10^{-7}$-$10^{-6}$ $\Omega$cm/K$^2$, estimated in $Fe_2VAl_{0.95}$ reminds one of the mass enhancement in the stoichiometric samples [6, 8]. To compare with the Kadowaki-Woods law, we should consider the small carrier concentration, $n$, in $Fe_2VAl$ and related alloys [7, 10]. In the Fermi liquid theory, $A/\gamma^2$ is proportional to $1/k_F^4$, where $\gamma$ and $k_F$ are the Sommerfeld constant in the specific heat and the Fermi wave vector, respectively [39]. Compared with heavy fermion systems with $n \approx 1$ (per rare earth or actinide element), it is expected that $A/\gamma^2$ is enhanced by a factor of $1/n^{4/3} = 2300\text{-}360$ in the case of $Fe_2VAl$, since $n = 0.003\text{-}0.012$ [7]. In this estimate, we assumed $k_F \sim n^{1/3}$ given by the Drude model (the free electron model) [40]. Therefore, $A/\gamma^2$ can be a measure of $n$, or in other words, the Fermi surface volume. In $Fe_2VAl_{0.95}$, $A/\gamma^2$ is $2\times10^{-3}$ $\Omega cmK^{-2}/(JK^{-2}mol^{-1})^2$ being a factor of 40 larger than that established as the Kadowaki-Woods ratio $\sim 1\times10^{-5}$ $\Omega cmK^{-2}/(JK^{-2}mol^{-1})^2$. Therefore, $n \sim 40^{-3/4} = 0.063$ per unit formula in $Fe_2VAl_{0.95}$. This value of $n$ is one order of magnitude larger than in $Fe_2VAl$ [7] and seems to be consistent with the value of the Hall coefficient of $Fe_2VAl_{0.95}$ [7, 12]. The observed value $\gamma \approx 12$ $mJmol^{-1}K^{-2}$ at $H = 0$ is rather large when we consider the carrier concentration, $n \approx 0.06$, and the value of normal metals, $\gamma \approx 1$ $mJmol^{-1}K^{-2}$. Consequently, the enhancement factor $m^*/m_e$ is $\sim 30$, where $m_e$ is the bare electron mass. Compared with the values $\gamma \approx 1.5\text{-}2.2$ $mJmol^{-1}K^{-2}$ and $n = 0.003\text{-}0.012$ [7] in non-magnetic $Fe_2VAl$, this large enhancement of $\gamma$ in ferromagnet $Fe_2VAl_{0.95}$ is surprising because $Fe_2VAl$ is located closer to the magnetic quantum critical point, $T_C \to 0$ [6], where the magnetic fluctuations should be more strongly enhanced. As $C/T$ below $T_C$ has an anomalous temperature and magnetic field dependence, we speculate that the enhancement mechanism of $\gamma$ distinctly differs from the electron-electron correlation mechanism realized in the heavy-fermion systems. Stoichiometric $Fe_2VAl$ is located at the vicinity of a ferromagnetic threshold composition, $\delta_c$. However, not only the Curie temperature, $T_C$, but also the paramagnetic moment, $\mu_p$, which is strongly dependent on the number of conduction carriers [7, 22-23] seem to disappear at $\delta = \delta_c$ in $Fe_2VAl_{1-\delta}$. In contrast, heavy fermion systems show only the magnetic transition temperature, $T_M$, going to zero while keeping a magnetic moment which has both a localized and an itinerant nature. In other words, the paramagnetic moment does not vanish around the critical point in heavy fermion systems.

## IV. Summary

Magnetic, transport and specific heat measurements have been carried out on the Heusler type compound $Fe_2VAl_{0.95}$. The data reveal a ferromagnetic transition takes place at $T_C = 33$ K, with the characteristics of an itinerant electron ferromagnet. With applying pressure, $T_C$ decreases with the pressure coefficient of $(1/T_C)(dT_C/dP) = -0.061$ $GPa^{-1}$. The Sommerfeld coefficient in the specific heat is enhanced, $\gamma \approx 12$ $mJmol^{-1}K^{-2}$, compared to 3 $mJmol^{-1}K^{-2}$ in $Fe_2VAl_{1.05}$ and 2 $mJmol^{-1}K^{-2}$ in stoichiometric $Fe_2VAl$. The temperature dependence of resistivity deviates from the characteristics of a Fermi liquid, $AT^2$, while under magnetic field the quadratic temperature dependence, $\rho(T) \sim T^2$, appears below a certain temperature, $T^*$, which increases with increasing magnetic field. At lower temperature the temperature variation of the specific heat shows a deviation from the $T$-linear dependence, especially, under magnetic field, which can be fitted by the sum of a $T$-linear and a $T^2$ terms. While the magnetism has a typical characteristics of a weak itinerant ferromagnet, the deviations from the Fermi liquid characteristics are quite different from those (*i.e.* non-Fermi liquid behavior) in a heavy fermion system and a nearly or a weak itinerant ferromagnet.

**Acknowledgement**

One of the authors (TN) appreciates valuable discussions with Dr. Matsubayashi (ISSP, Univ.

Tokyo) and thanks NWO (Dutch Organisation for Scientific Research) for a visitor grant.

**Figures and Captions**

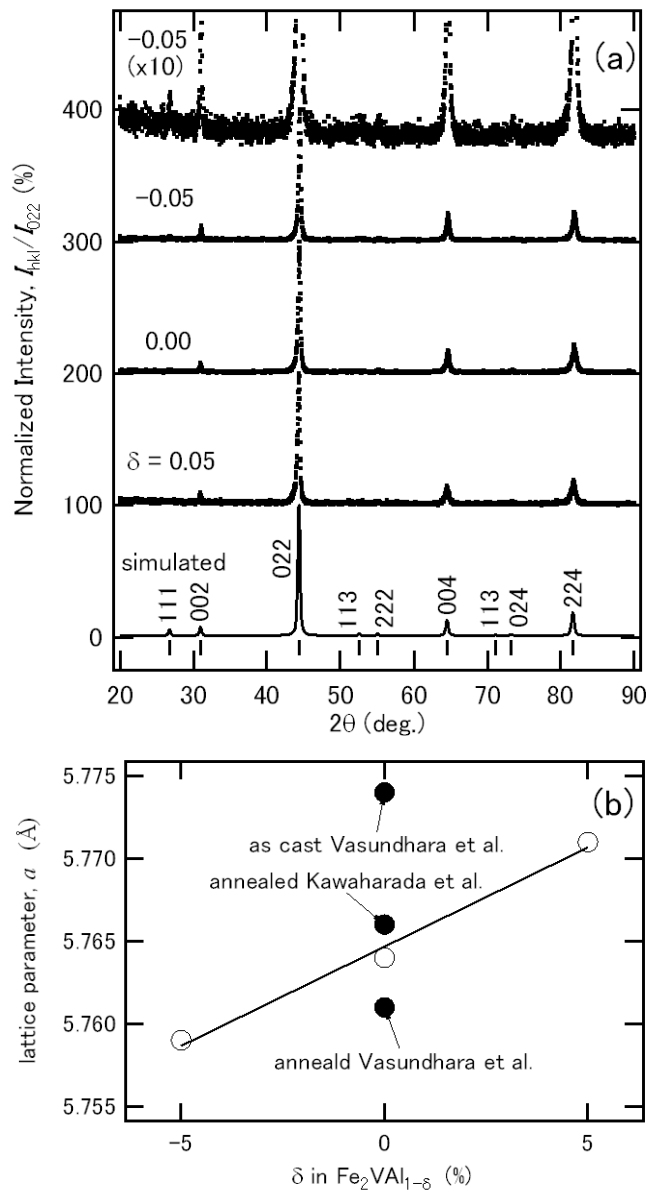

Fig. 1 (a) X-ray diffraction patterns of $Fe_2VAl_{1-\delta}$ at $\delta$ = -0.05, 0.00, 0.05 and a simulated one for $Fe_2VAl$ with the Heusler structure. The top XRD pattern of $\delta$ = -0.05 is magnified ten times vertically. Ticks and numbers being along the simulated pattern indicate the position of the Bragg reflections and corresponding Miller indices, respectively, for the Heusler structure.

(b) Lattice parameters obtained in this work as a function of $\delta$ (open circle). Solid line is a guide for eyes. Solid circles at $\delta$ = 0 represent reported values in Refs. 18 and 20.

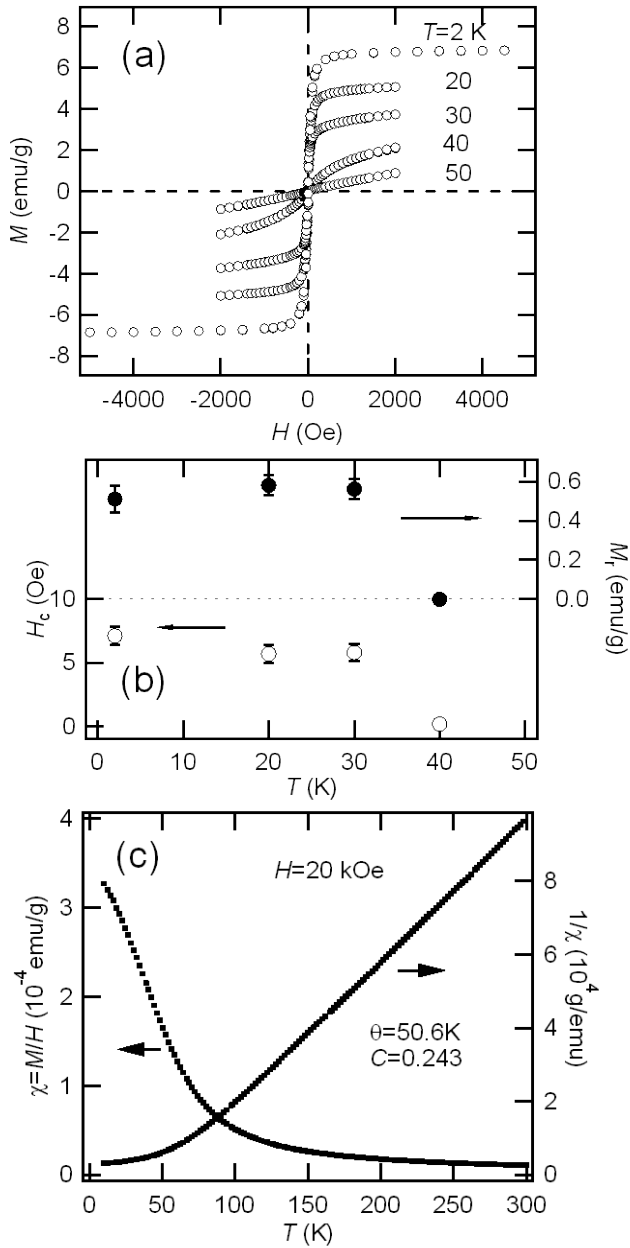

Fig. 2 (a) Magnetization *versus* field (*M-H*) of $Fe_2VAl_{0.95}$ at fixed temperatures as indicated.
(b) Coersive force, $H_c$, and remanent magnetization, $M_r$, as a function of temperature.
(c) Magnetic susceptibility, $\chi(T) = M/H$, and reciprocal susceptibility, $1/\chi(T)$, as a function of temperature measured in a fixed field *H*=20 kOe. The susceptibility is expressed in units per gram.

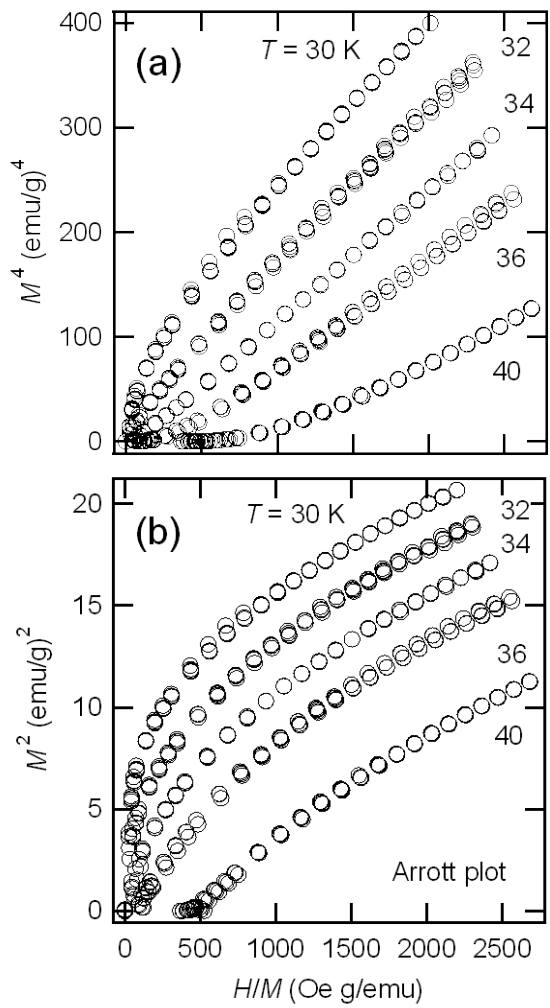

Fig. 3 Magnetization of $Fe_2VAl_{0.95}$ measured at fixed temperatures as indicated, plotted as
  (a) $M^4$ *versus* $H/M$
  (b) $M^2$ *versus* $H/M$ (Arrott plot).

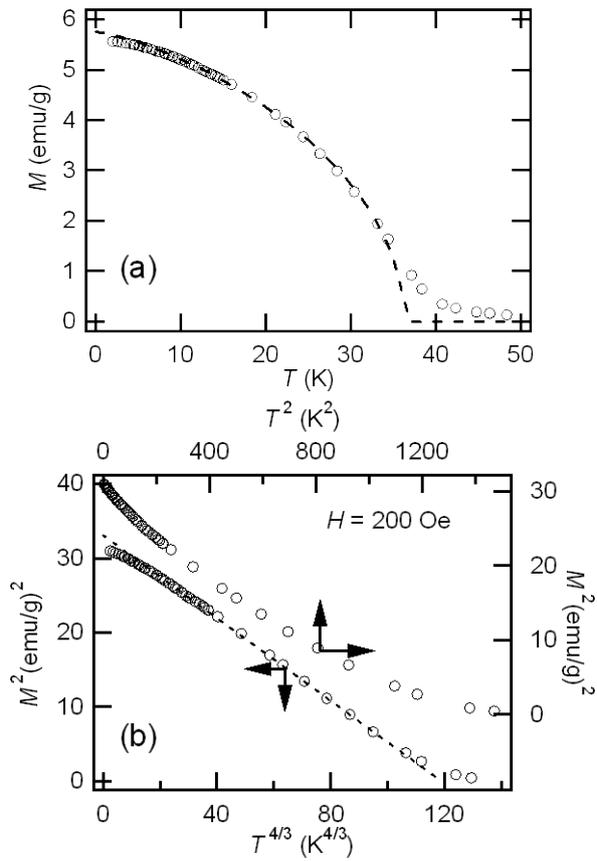

Fig. 4 (a) Magnetization in a field $H$ = 200 Oe as a function of temperature for $T$ < 50K.
(b) $M^2$ - $T^{4/3}$ and $M^2$ - $T^2$ plots measured in a field $H$ = 200 Oe. The dashed lines are calculated by using $M^2(T) = M(0)(1-(T/T_C)^{4/3})^{1/2}$ with $M(0)$ = 5.7 emu/g and $T_C$ = 36 K obtained by a least-square fit.

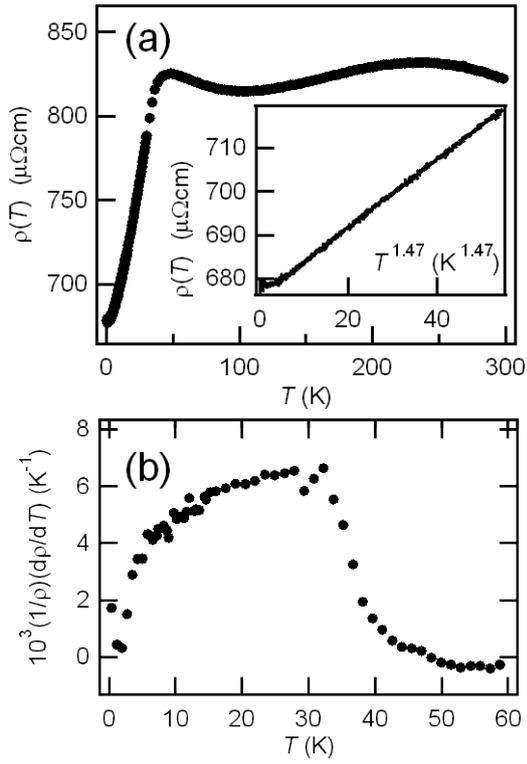

Fig. 5 (a) Resistivity ρ(*T*) as a function of temperature at *H*=0. Inset: Resistivity below $T_C$ plotted as a function of $T^{1.47}$.

(b) Temperature variation of (1/ρ)(dρ/d*T*). The slope change at *T* = 33 K signals $T_C$.

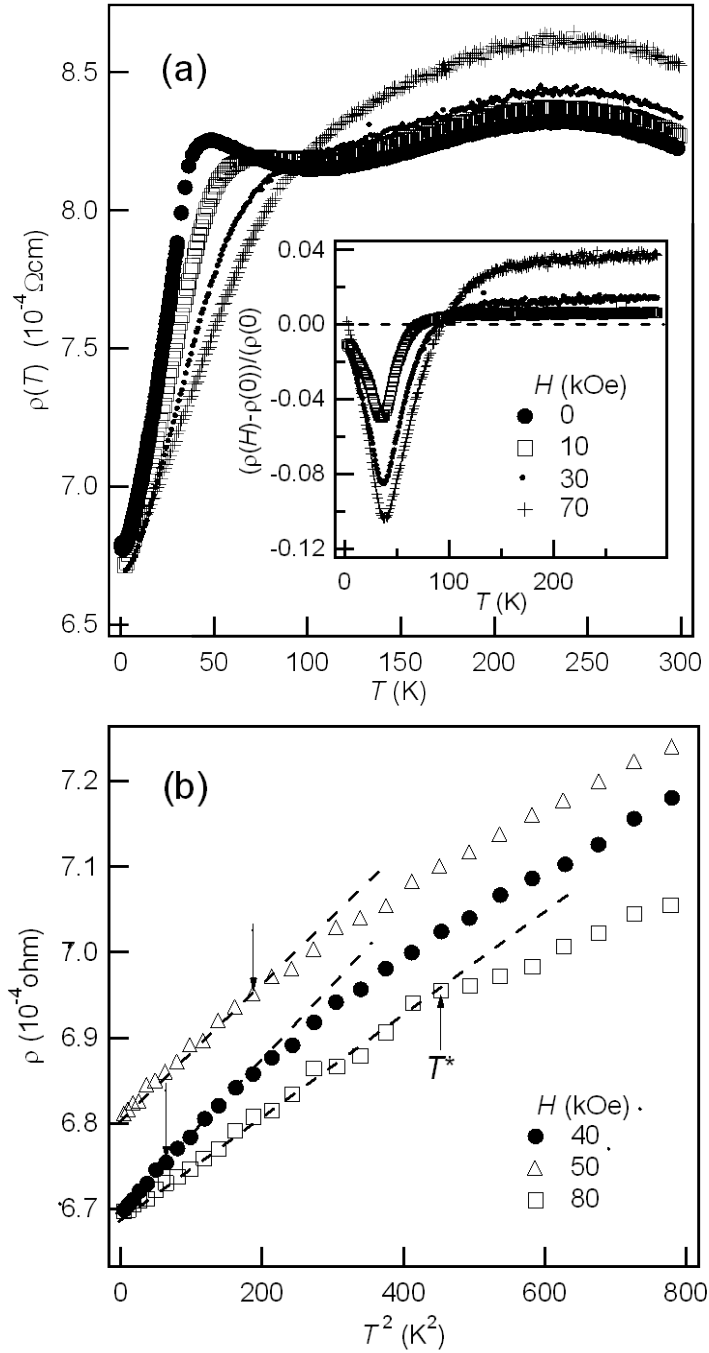

Fig. 6 (a) Resistivity *versus* temperature of $Fe_2VAl_{0.95}$ in magnetic fields up to 70 kOe as indicated. Inset: Temperature dependence of the magnetoresistance $(\rho(H) - \rho(0))/\rho(0)$ at $H = 10$, 30 and 70 kOe.

(b) Resistivity as a function of $T^2$ at $H = 40$, 60 and 80 kOe. The $T^2$ behavior (dashed lines) is obeyed till a temperature $T^*$ indicated by vertical arrows.

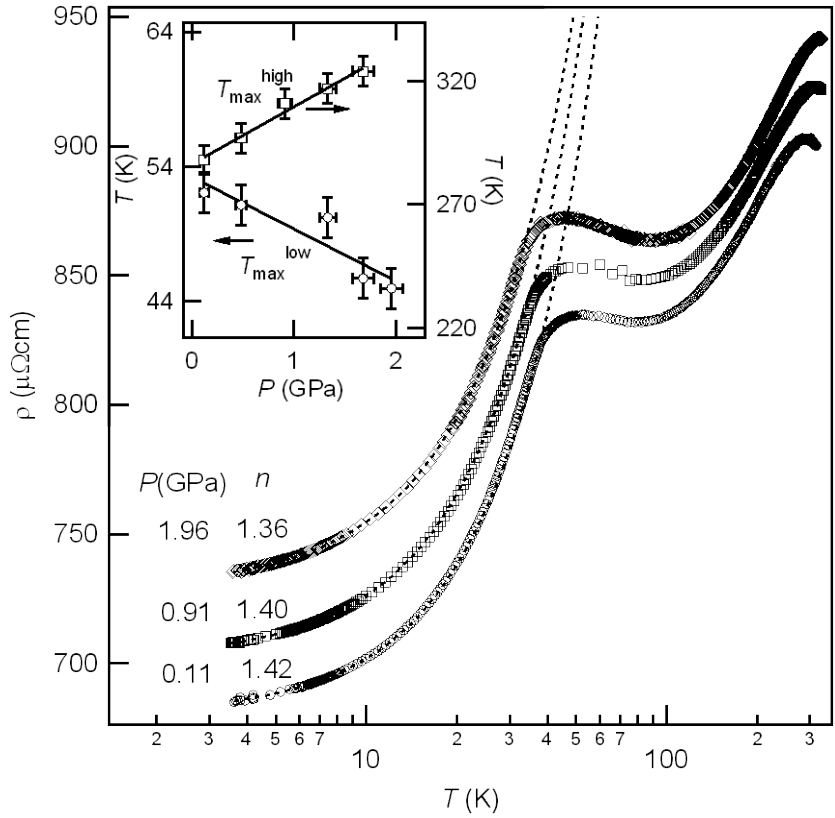

Fig. 7 Resistivity *versus* temperature of $Fe_2VAl_{0.95}$ at various pressures. Dashed line shows a power low dependence of $\rho(T) \sim AT^n$. The exponent, *n*, was obtained by fitting $\rho(T)$ to $\rho(T) = \rho_0 + AT^n$ below 11 K. Applied pressure, *P*, and obtained exponent, *n*, are indicated on the respective $\rho(T)$ curve. Inset: Pressure dependence of the maximum temperatures, $T_{max}^{low}$ and $T_{max}^{high}$. Solid lines are guides for the eyes.

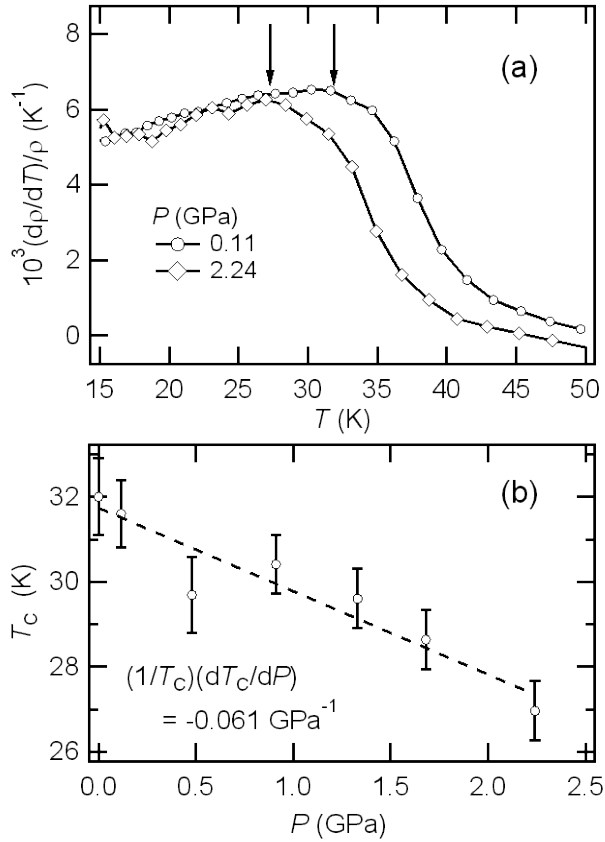

Fig. 8 (a) Temperature dependence of $(1/\rho)(d\rho/dT)$ at various pressures. Arrow indicates Curie point defined as the maximum point of $(1/\rho)(d\rho/dT)$.

(b) Pressure variation of $T_C$. Dashed line represents a fitting curve of $T_C$ versus $P$ obtained by the linear least squares method.

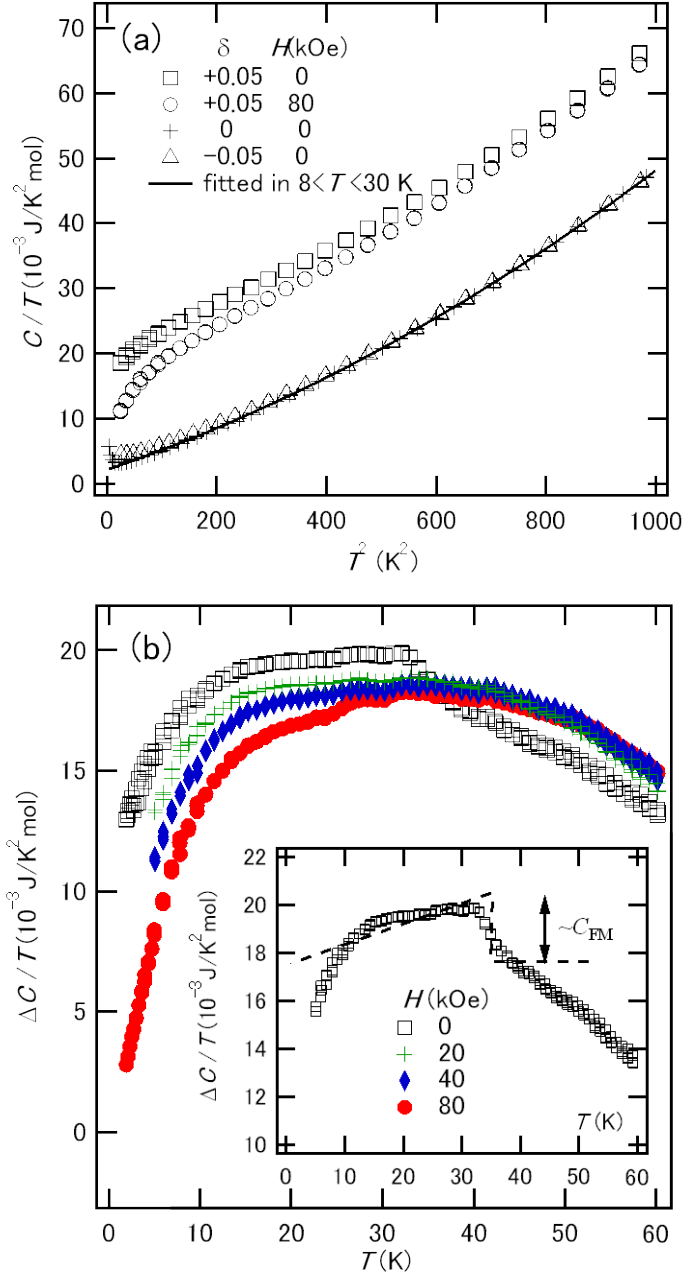

Fig. 9 (a) Specific heat divided by temperature $C/T$ plotted as a function of $T^2$ of Fe$_2$VAl$_{0.95}$ ($\delta$= +0.05), Fe$_2$VAl ($\delta$= 0) and Fe$_2$VAl$_{1.05}$ ($\delta$= -0.05) in zero and applied magnetic field. The solid line represent the fit results in the temperature region 8 K < $T$ < 30 K (see text).

(b) Temperature dependence of $\Delta C/T$ defined by $(C(\delta= +0.05)-C(\delta= 0))/T$ at fields of 20, 40 and 80 kOe as indicated. Inset: Expanded view of $\Delta C/T$ around the Curie temperature at zero field. The dashed line represents an idealized ferromagnetic contribution, $C_{FM}$.